\documentclass[preprint,aps,pre,showpacs,notitlepage,superscriptaddress,longbibliography,floatfix]{revtex4-2}

\usepackage{epsfig,graphics,amssymb,amsmath,subeqnarray,color,bm,bbm}
\usepackage[toc,page]{appendix}
\usepackage{mathtools}
\usepackage{xcolor}
\usepackage{float}
\usepackage{graphicx}
\usepackage{subfigure}
\usepackage[colorlinks=true,allcolors=blue]{hyperref}
\usepackage[toc]{appendix}

\usepackage{booktabs}
\usepackage{parskip}
\usepackage{comment}
\usepackage{esdiff}
\usepackage[arrowdel]{physics}
\usepackage{verbatim}
\usepackage{array}

\begin{document}
\title{The motility-matrix production switch in {\it Bacillus subtilis} -- a modeling perspective}
\author{Simon Dannenberg}
\author{Jonas Penning}
\author{Alexander Simm}
\author{Stefan Klumpp}
\email{Stefan.klumpp@phys.uni-goettingen.de}
\affiliation{Institute for the Dynamics of Complex Systems, University of G{\"o}ttingen, Friedrich-Hund-Platz 1, 37077 G{\"o}ttingen, Germany.}
\date{\today}


\begin{abstract}

Phenotype switching can be triggered by external stimuli and by intrinsic stochasticity. Here, we focus on the motility-matrix production switch in \emph{Bacillus subtilis}. We use modeling to describe the SinR-SlrR bistable switch its regulation by SinI, and to distinguish different sources of stochasticity. Our simulations indicate that intrinsic fluctuations in the synthesis of SinI are insufficient to drive spontaneous switching and suggest that switching is triggered by upstream noise from the Spo0A phosphorelay.

\end{abstract}

\maketitle

\section{Introduction}

Cell fate decisions in which cells switch from one phenotype to another adapting their morphology, metabolism, and gene expression or signaling programs are common from microbial stress responses to developmental pathways in higher organisms \cite{veening2008bistability,ferrell2012bistability,ferrell1998biochemical,greenwald1998lin}. They raise multiple questions including how much these switches are determined by external triggers such as changes in their environment, what role stochastic noise plays in switching, and whether the change in phenotype is only transient or irreversible. Correspondingly, the underlying biochemical and genetic circuitry may display a variety of different dynamic behaviors including bistability and excitability \cite{veening2008bistability, dubnau2006bistability,suel2006excitable}.
One of the most drastic changes in the behavior of bacterial cells is the transition from single-celled to multicellular lifestyles  \cite{lyons2015evolution,shapiro1988bacteria}. 
From a conceptual point of view, multicellular behavior can manifest itself at two levels: On one level, cells may differentiate into specialized cell types that perform tasks that provide a fitness benefit to the community rather than to the individual cell, typically either by division of labor between different cell types or through bet hedging. Examples include metabolic differentiation  and the formation of spores and persister cells \cite{flores2010compartmentalized,balaban2011persistence,mutlu2018phenotypic,veening2008bistability}. On a second level, cells can form tissue-like multicellular communities that are mechanically coupled to each other via cell-cell contacts or via a matrix \cite{hall2004bacterial,arnaouteli2021bacillus} and form complex three-dimensional structures that provide both protection against outside stresses \cite{hall2004bacterial} and allow supply of nutrients \cite{wilking2013liquid}. In many cases, however, differentiation and mechanical coupling occur together. For example, only a subset of cells might produce matrix. Independent of that, both dimensions of multicellularity are based on phenotypic switches that induce matrix formation and/or cellular differentiation. 



Biofilm formation in the model organism {\it Bacillus subtilis} is a prime example in which bacteria form a multicellular community with both features: The cells form a three-dimensional structure via a matrix that consists of exopolysaccharides, protein fibres and extracellular DNA and that encapsulates the cells, coupling them mechanically. This structure is formed via differentiation into different cell types including matrix producers, motile cells and spores \cite{arnaouteli2021bacillus,Vlamakis}.
The formation of a biofilm in \textit{Bacillus subtillis} is initiated by  a switch between the planktonic motile state and the matrix-producing state that  typically occurs only in a subset of cells \cite{Chai2008}. The switch is controlled by the proteins SinR and SlrR (Figure \ref{network_bistabiltiy}). SinR represses genes required for matrix formation (the exopolysaccharide production operon {\it epsA-O} and the {\it tasA} gene encoding the main protein component of the matrix) as well as the gene encoding SlrR \cite{Kearns2005,Chu.2006}. Vice versa, SlrR inactivates SinR by forming a SinR-SlrR complex \cite{Chai2010}. The latter complex also represses genes required for motility (specifically the {\it hag} gene that encodes flagellin) and for cell separation \cite{Chai2010,kobayashi2008slrr}. Another key regulator is SinI, which also sequesters SinR in a complex and inactivates it \cite{Bai1993, newinsights}. SinI in turn is activated by the phosphorylated form of the global regulator Spo0A \cite{Fujita.2005}, which in turn is activated by a phosphorelay \cite{Burbulys1991} and controls both biofilm formation and sporulation \cite{lopez2010extracellular}.



The transition to matrix formation is often induced in response to nutrient depletion. When studying the dynamics under nutrient depletion, the dynamics is influenced by direct regulation and indirect effects of the cell's physiological state \cite{Igoshin2023}. However, using a constant environment provides the advantage of decoupling the dynamics of the genetic circuit underlying the phenotype switch from global control through growth or the cell's physiological state in general \cite{Klumpp2009}. Such observations of individual cells expressing fluorescence reporters and their lineages can be achieved in microfluidic devices such as the mother machine \cite{wang2010robust}. This allows one to study the the dynamics of the motile and the matrix-producing state in individual cells and the stability of these phenotypes over many generations \cite{Norman,Paulsson2019,kampf2018selective}. Remarkably, in this setup stochastic transitions between the motile state and matrix formation are observed even under nutrient-rich conditions \cite{Norman}. Here, they found the motility state of \textit{B. subtilis} to be very stable with an exponentially distributed lifetime with an average of 81 generations \cite{Norman} and the transition to the matrix-producing state is  stochastic and memoryless. The matrix-producing state is shorter-lived (7.6 generations) with a peaked rather than exponential distribution and has been interpreted as exhibiting memory, indicating a stereotypical dynamic program of transient inducting of the matrix producing state \cite{Norman}. This proposes the question how the architecture of the underlying dynamics itself can provide an internal switch between these two states that is independent of external stimuli.

The regulatory network depicted in Figure \ref{network_bistabiltiy}A suggests two possible mechanisms that could induce the transition to matrix production and both have been proposed in the literature: Pulses of SinI in a subpopulation of cells trigger the induction of matrix producing genes and the expression in only a subpopulation appears to depend on the heterogeneous activation of Spo0A \cite{Chai2008}. An alternative  mechanism is based on the stochastic competition between SinI and SinR and is supported by the observation that stochastic switching can be seen in a SinI-SinR circuit decoupled from Spo0A \cite{Paulsson2019}. 

Here, we use a mathematical model for the phenotype switch to ask what is the source of the stochasticity driving these transitions. We first set up a deterministic model to study the dynamics of the switch upon a deterministic SinI pulse. We then use several variants of stochastic models -- for the SinR-SlrR switch, the SinR-SinI competition and the coupling to the phosphorelay, for the latter building on previous work in the context of sporulation \cite{Igoshin2} -- to test different sources for the stochastic switch to matrix production. Our results suggest that neither the spontaneous fluctuations in the mutual repression of SinR and SlrR nor those in the stochastic competition between SinI and SinR expression are likely the source. Instead,  our results suggest that the dominant source of fluctuations is upstream of the phenotype switch and the noise required for switching is transmitted via the Spo0A phosphorelay.

\section{results}
\subsection{Bistability without SinI}

\begin{figure}[tb]
    \centering
    \includegraphics[width = \textwidth]{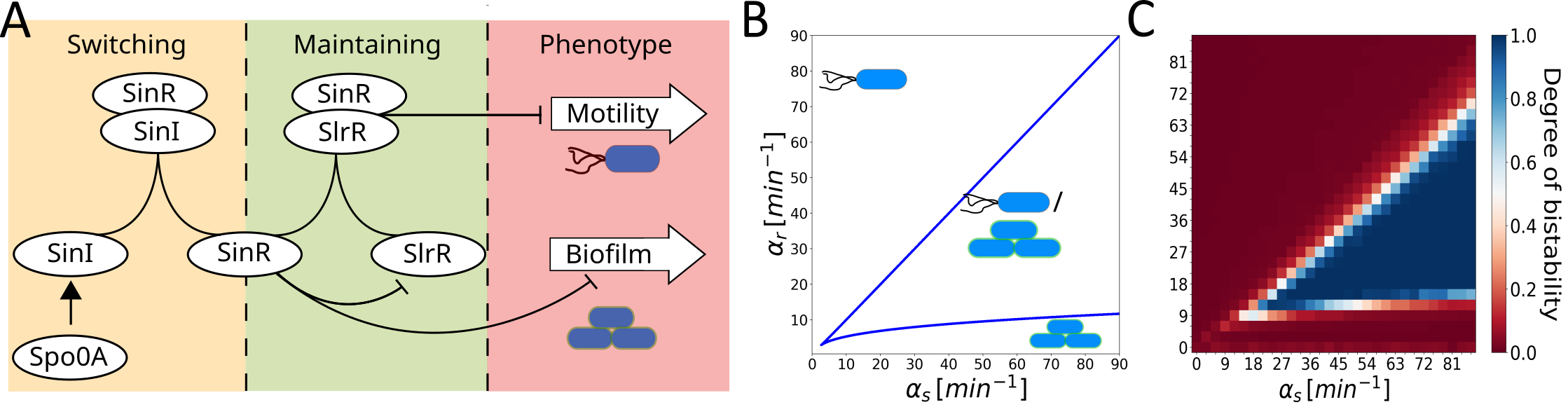} 
    \caption{Bistability in the SinI-SinR-SlrR regulatory network: \textbf{A} Schematic overview of the SinR-SlrR-SinI regulatory network for the phenotype switch between motility and biofilm formation/matrix production. Arrows indicate activation, arrows with bars repression and joining lines complex formation. 
    \textbf{B} State diagram of the network showing the phenotype as a function of the synthesis rates of SinR and SlrR, $\alpha_r$ and $\alpha_s$ respectively. For small $\alpha_r$, the network displays the matrix-producing phenotype, for large $\alpha_r$, the motile phenotype. The sketches indicate the corresponding phenotypes of the steady state solutions. For intermediate values, between the two blue lines, the system is bistable with coexistence of the two phenotypes. \textbf{C} The degree of bistability, determined from stochastic simulations, as a function of the two synthesis rates. A value of $1$ (blue) indicates that two stable solutions exist and the system is bistable, while a value of $0$ (red) corresponds to only one stable state.  }
    \label{network_bistabiltiy}
\end{figure}

With the goal of illuminating the roles of the relevant proteins and interactions, we first simulated the SinR-SlrR switch with a deterministic model. To that end, we expressed the control of the  switch as depicted by the regulatory network in Figure \ref{network_bistabiltiy}A by rate equations (see Methods for a detailed description). This allows us to  monitor the time development of the protein concentrations and thereby the dynamics of the system. Importantly, we can study isolated modules of the network insulated from additional influences. For example, while SlrR is essential for matrix formation, the initiation of matrix formation is unchanged in SlrR mutants \cite{Norman}, supporting such modularity  in the network with the SinR-SinI interaction controlling the initiation of biofilm formation and the SinR-SlrR switch controlling the maintenance of the matrix production and motility states.

Therefore, we first studied the SinR/SlrR switch without its control by SinI
  (center and right part of Figure \ref{network_bistabiltiy}A). This module consists of  SlrR, SinR and the SlrR-SinR complex, with concentrations, $s$, $r$ and $c$, respectively. Their interactions are given by the repression of {\it slrR} by SinR and the irreversible complex formation of the two molecules, which inactivates SinR and thus relieves repression by SinR. The dynamics of this module is given by
\begin{equation}
    \begin{split}
        \Dot{s}  &= \alpha_s R(r) - k_+ s r - \beta s \\
        \Dot{r}  &= \alpha_r  - k_+ s r - \beta r \\
        \Dot{c}  &= k_+ s r - \beta c \\
    \end{split}
\label{eqwithoutsinI_short}
\end{equation}
Here $\alpha_s$ and $\alpha_r$ are the synthesis rates of SlrR and SinR, respectively, $\beta$ is the dilution rate due to cell growth. The synthesis of SlrR is modulated by a Hill function $R(r)$ to describe the repression. $k_+$ is the complex formation rate and complex formation is taken to be irreversible. The two phenotypes are identified via reporter proteins which are located downstream of the network and repressed by SinR and SlrR, respectively (matrix reporter and motility reporter). 
A more detailed description of the equations and estimates of the parameter values are given in the Methods section. 

\begin{figure}[!h]
    \centering
    \includegraphics[scale = 0.5]{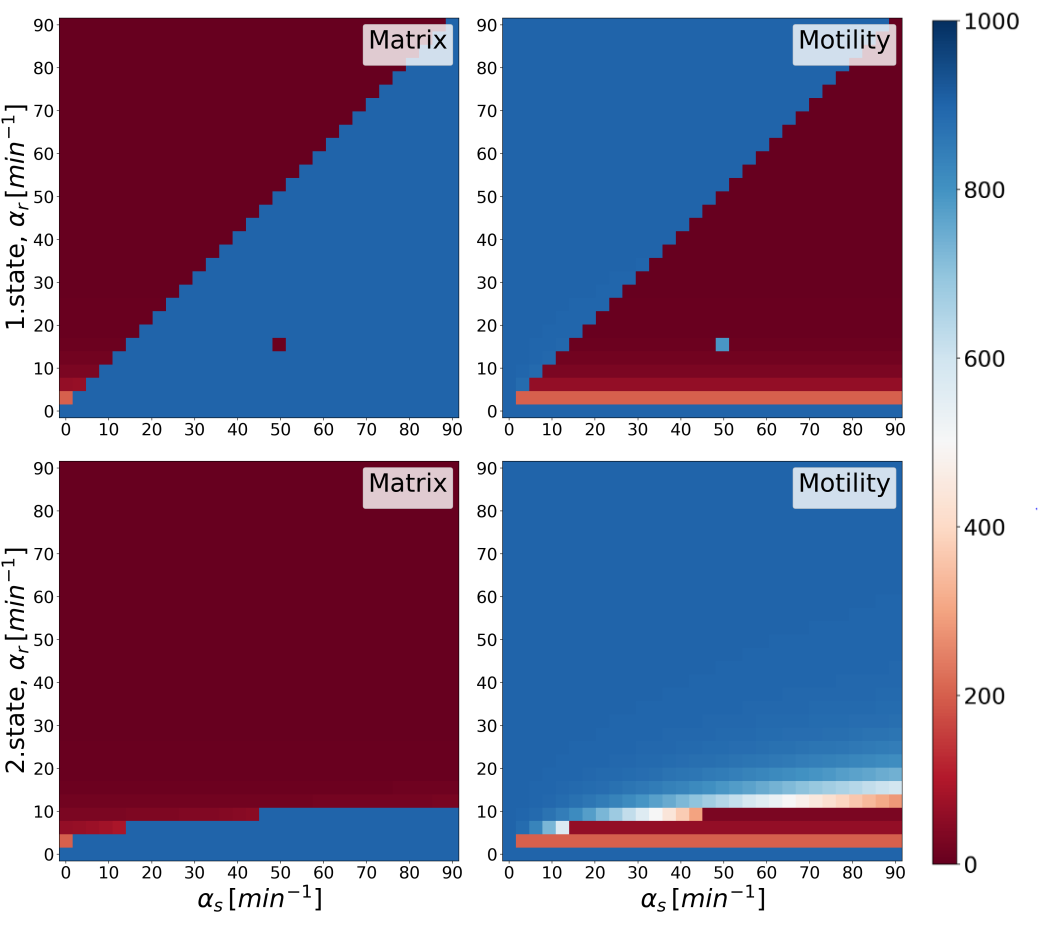}
    \caption{Concentrations of the reporter proteins in the stable steady states. The column depicts either the matrix or the motility reporter. The row is used to depict the two different options in the bistable region.}
    \label{Steady_state_concentrations}
\end{figure}

The coexistence of multiple phenotypes, here matrix formation and motility, is often based on an underlying bistablity in the dynamics of the genetic network, which enables two steady state solutions to emerge under the same conditions. To see whether the SinR/SlrR module of the network can display  such bistability on its own, we investigated if parameter regions  of $\alpha_r$ and $\alpha_s$ exist, in which multiple steady state solutions are present. To that end, we determined bifurcations points of the dynamics (see Methods).  
These are  plotted in Figure \ref{network_bistabiltiy}B, varying the synthesis rates. The concentrations of the reporters in the corresponding steady states are plotted in figure \ref{Steady_state_concentrations} as a heat map. This analysis shows an approximately triangular area, in which two steady state solutions co-exist. These correspond to high expression of biofilm genes and low expression of the motility genes and vice versa. Below the triangular bistable area, a monostable biofilm solution exists, while above it a monostable motility region can be found.
Subsequently, we investigated the stability of the individual states of the  bistable solutions to see whether spontaneous transitions between them occur. To this end, we simulated a stochastic version of the model above using the Gillespie algorithm \cite{Gillespie} and considering bursty production of proteins (for a detailed description, see Methods). To determine the robustness of these states against fluctuations, we analysed how long they are maintained on average. For a quantitative assessment of bistability, we evaluate the degree of bistability $B$ by calculating the mean fraction of time a trajectory stays in its initial state, starting from the matrix and motility state. $B$ is calculated as  $B = {f}_{Mo} + {f}_{Ma} - 1$, where $f_{Ma,Mo}$ describes the average fraction of time spent in the motility or matrix state during a simulation run started in that respective state. For each average 100 simulations are performed for 50 hours. Longer simulations showed similar behaviour (not shown).
With this definition, $B$ is 1 if the system stays in either initial state and is thus bistable, but approximately 0 if only one of the states is stable.
The resulting values for $B$ are plotted in a heat map in Figure \ref{network_bistabiltiy}C. Most of the bistable region does not show any transitions between the solutions and can therefore be considered to be stable over long times. Only near the boundaries of the bistable area, transitions between the two states are observed, resulting in $B$ values between 0 and 1.
These results show that the interaction of SlrR and SinR results in  robust bistability over a large parameter range and therefore provides the basis for a switch between two stable phenotypes with rare spontaneous transitions.

\subsection{Effect of SinI}
\begin{figure}[tb]
    \centering
    \includegraphics[scale = 0.20]{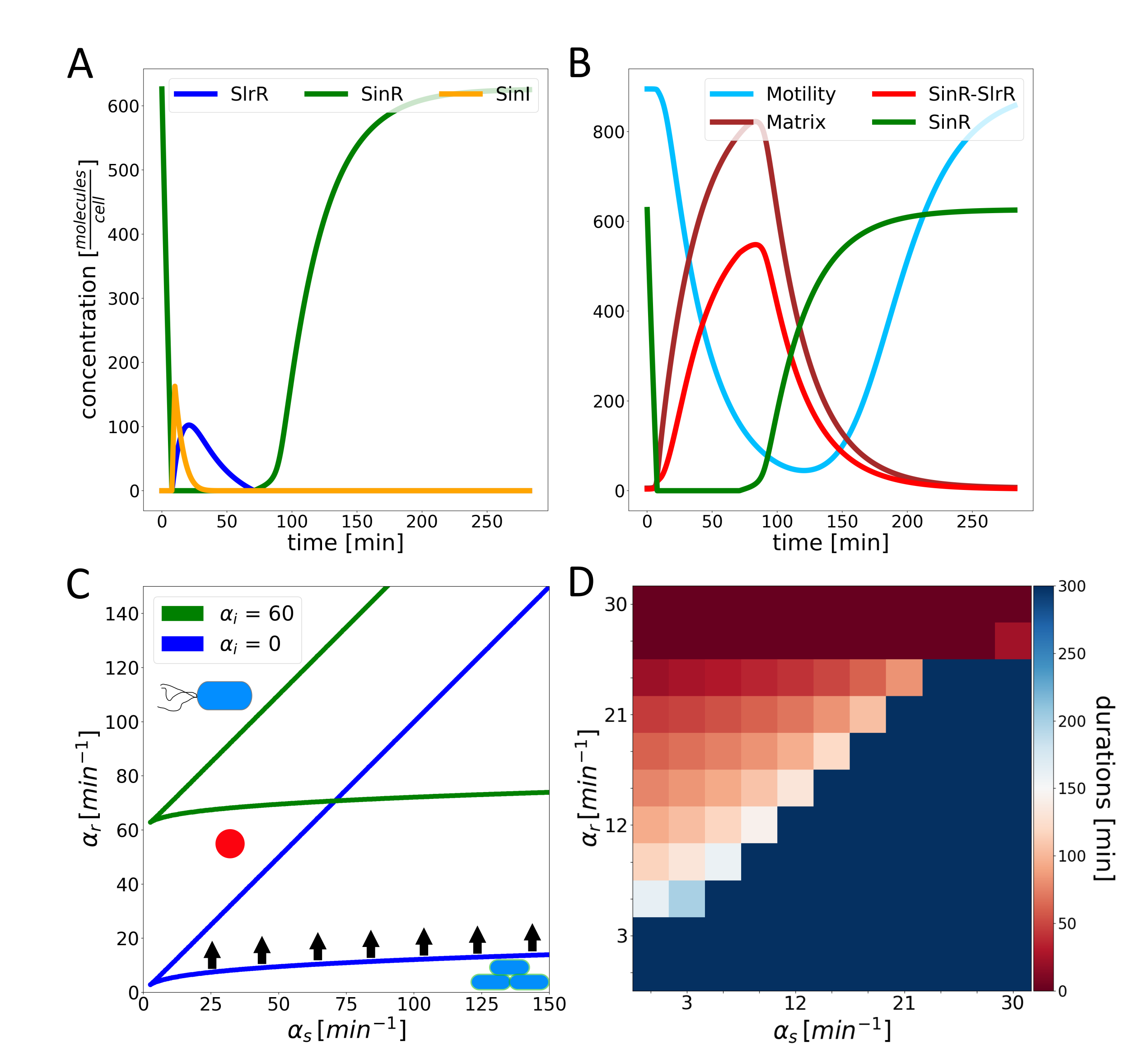}
    \caption{Switch to matrix production upon a SinI pulse: \textbf{A, B}  Time evolution of the concentrations of the switch proteins (SinR, SlrR, SinI, SinR-SlrR complex) and motility and matrix reporter genes following a pulse of SinI synthesis as obtained from the deterministic model (rate $\alpha_I = 90 \, \frac{\text{molecules}}{\text{min}}$ for 10 minutes). Parameter values of $\alpha_r = 21$ min$^{-1}$ and $\alpha_s = 20$ min$^{-1}$ have been used.  \textbf{C} State diagram showing the stable phenotypes as a function of the synthesis rates $\alpha_r$ and $\alpha_s$ of SinR and SlrR, respectively, for scenarios with and without SinI synthesis (bistable region indicated by the green and blue lines, respectively). SinI synthesis shifts the area of bistability towards larger $\alpha_r$ as indicated by the black arrows. The red dot marks a parameter combination for which  the network displays the motility state in absence of SinI (above the upper blue line) and is moved to the matrix-producing state (below the lower green line) by turning on the SinI synthesis. \textbf{D} The duration of the matrix-producing periods after such a SinI pulse for varying values of $\alpha_r$ and $\alpha_s$. In the dark blue region the system does not return to the motility state before the end of the simulation. In the dark red region the SinI pulse does not lift the SinR repression. Note the different parameter ranges of \textbf{C} and \textbf{D}}
    \label{involving_SinI}
\end{figure}

To elucidate the basis of switching, i.e., of transitions between the two phenotypes, we considered an extended network that includes SinI in addition to SinR and SlrR (see Methods). The extended model includes the synthesis and degradation of SinI and, most importantly, its complex formation with SinR, described in analogy to the SinR-SlrR complex formation above. Through complex formation, SinI can effectively titrate SinR and thereby lift the repression of the {\it slrR} gene and of the reporter for matrix production.

The switching dynamics can be illustrated by looking at a system which is initially in the motile state but gets triggered by a short SinI pulse (Figure \ref{involving_SinI} A,B). We prepared the system in the steady state of the monostable motility region by choosing $\alpha_r = 21$ min$^{-1}$ 
and $\alpha_s = 20$ min$^{-1}$. 
Subsequently, the SinI synthesis is turned on by setting $\alpha_I = 90$ min$^{-1}$ for 10 minutes, mimicking a SinI pulse as observed \cite{Chai2008}. With this, the newly synthesized SinI molecules  titrate the  SinR concentration and  allow for the relief of the SinR repression so that SlrR and the biofilm reporter are produced. At the same time the SlrR-SinR complex is formed and the motility reporter gets repressed. These two effects lead to a switch in the dominant reporter concentration. After the SinI synthesis has stopped, the synthesis of SlrR continues until  the SinR concentration recovers and represses SlrR again causing the system to fall back into its initial state. \\

The role of SinI can be illustrated by the change of bifurcation points if SinI synthesis is turned on. We calculated the resulting bistability regions in the parameter space without and with SinI (Figure \ref{involving_SinI}C, blue and green triangular regions for $\alpha_I=0$ and $\alpha_I=60$ min$^{-1}$, respectively). The triangular bistability region is shifted upwards by the introduction of SinI. This behaviour can be explained by the strong irreversible binding of SinI and SinR, which titrates out pairs of of free SinR and SinI molecules. This titration effectively reduces the synthesis rate of SinR  to an effective rate  $\alpha_{r, {\rm eff}} = \alpha_r - \alpha_I$ in equation \ref{eqwithoutsinI_short}, resulting in the observed shift.
Importantly, the addition of SinI enables a switching mechanism between the two phenotypes. We illustrate this by considering a parameter combination (synthesis rates $\alpha_r$ and $\alpha_s$) which in the absence of SinI is located in the monostable motile region (indicated by the red dot in Figure \ref{involving_SinI}C). When the synthesis of SinI is induced, the bistability region shifts upwards (indicated by the black arrows) and the point in parameter space (red dot) is now located in the monostable region corresponding to the biofilm state. Thus, the phenotype is switched by the induction of SinR. 

At last we also  quantified this switching mechanism by performing a systematic analysis of the duration of the transient phases of matrix expression after turning on SinI for a short period of time. To this end, such SinI pulses were applied for different combinations of $\alpha_r$ and $\alpha_s$ and the time period in which the matrix production reporter is dominant was observed. The resulting times are plotted as a heat map in Figure \ref{involving_SinI}D. This analysis shows that the time spent in the matrix-producing state is rather sensitive to the exact combination of $\alpha_r$ and $\alpha_s$ and there are two qualitatively different regimes. For parameters where $\alpha_r < \alpha_s$, the system does not exit the matrix-producing state again, which makes the switch permanent. If one now increases $\alpha_r$, so that $\alpha_r > \alpha_s$, the switch become transient and the longest periods of matrix production can be achieved if $\alpha_s \lesssim \alpha_r$. Here the time in which SlrR can prevent its titration of SinR  gets longer even if the SinI synthesis is stopped. Further, larger lifetimes are reached for lower values of $\alpha_r$ due to the higher buffer in SinI concentration. This allows the titration of SinR for even longer periods. 
The sensitivity to the difference in $\alpha_r$ and $\alpha_s$ can also be seen in an analytical approximation  
for the time in which SinR is not present after a SinI pulse  
(see Supporting Information).
In summary, our analysis shows that  SinI enables a mechanism for switching from motility to the matrix-producing state. The dynamics of the switch is sensitive to the exact values of the synthesis rates of SinR and SlrR. Thus, small changes in these rates can either lead to a stable matrix-producing phenotype or to different durations of transient expression of matrix genes.

\subsection{Minimal SinR-SinI stochastic competition model}


We have seen above that a pulse of SinI expression can induce switching  to the matrix-producing state, either transiently or permanently, depending on the synthesis rates of SinR and SlrR. Spontaneous switching in the absence of SinI was very rare. We thus ask next whether fluctuations in SinI can trigger switching. This has indeed been proposed in a recent study \cite{Paulsson2019}, where stochastic competition between SinI and SinR was proposed as the core element of the switch. Following the model of Lord et al. \cite{Paulsson2019}, in which bursting is included as well, we next consider a minimal model for stochastic competition between SinI and SinR, not including SlrR and considering only one reporter protein that indicates matrix production. The matrix production and motility states are then distinguished by whether the concentration of that reporter exceeds a certain threshold.

\begin{figure}[tb]
\centering
\subfigure{\includegraphics[width=1\textwidth]{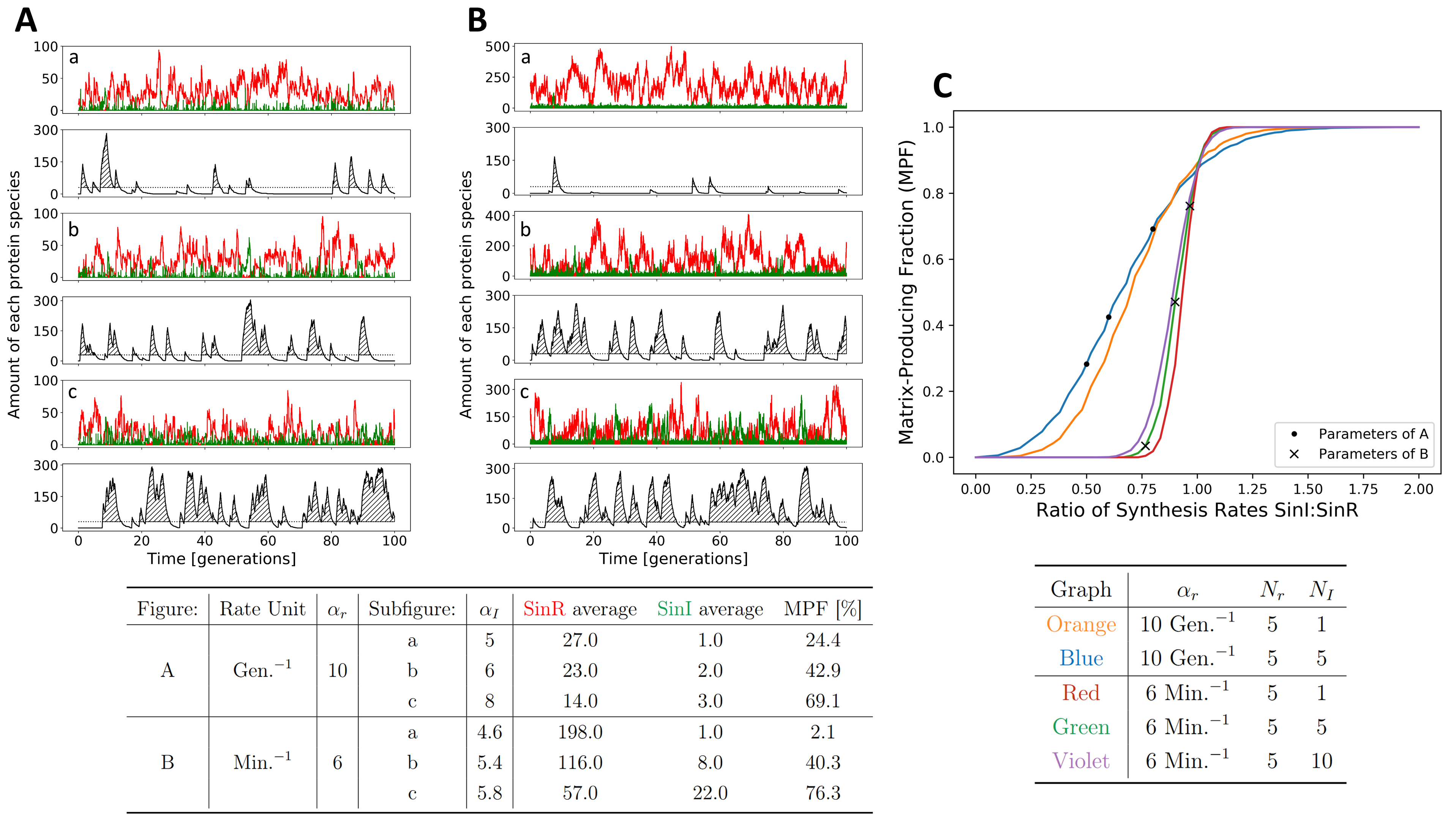}}
\caption{Stochastic competition of SinI and SinR: \textbf{A}, \textbf{B} Stochastic time evolution of the concentrations of SinR (red), SinI (green) and the matrix reporter protein (black) for different parameter combinations as indicated in the table below of the plots (parameters in \textbf{A} corresponds to those from Ref.~\cite{Paulsson2019}, those in \textbf{B} are estimated based the protein abundance from Ref.~\cite{Chai.2009}). When the reporter protein exceeds the threshold indicated by the dotted line (shaded area under the curve), the system is considered to be in the matrix-producing state. \textbf{C} fraction of cells in the matrix-producing state for the two parameter sets from \textbf{A} and \textbf{B}. Different burst sizes for SinI ($N_I$)  are used while the burst size for SinR is fixed to ($N_r=5$), such that the ratio of the effective synthesis rates $N_I\alpha_I/N_r\alpha_r$ varies between 0 and 2. For $N_I = 5$ (blue and green curve), symbols indicate the parameter combinations for which trajectories are shown in \textbf{A} and \textbf{B}. 
}
\label{fig:Jonas_1}
\end{figure}

To see whether fluctuations in the protein concentration can induce switching via a stochastic competition between SinI and SinR we simulated the dynamics of this minimal system. Trajectories of SinR, SinI and the reporter protein for matrix production are plotted in Figure \ref{fig:Jonas_1}A and B (red, green and black curves, respectively), varying the synthesis rates. 
In Figure \ref{fig:Jonas_1}A, the synthesis rate of SinR, $\alpha_r$, is chosen according to ref.~\cite{Paulsson2019}, resulting in relatively small numbers of SinR molecules, while in Figure \ref{fig:Jonas_1}B, $\alpha_r$ is adjusted such that the measured protein copy number of $\sim 400$ per cell \cite{Chai.2009} is matched (note the different scales on ther axes used in Figure \ref{fig:Jonas_1}A and B). The synthesis rate of SinI, $\alpha_I$ is increased from top to bottom.  
Both parameter sets show stochastic competition: the system switches between phases with and without reporter expression; in phases of high SinI expression, the SinR concentration is reduced and repression of the reporter is relieved. In both cases, the matrix production reporter is predominantly expressed for large $\alpha_I$ and predominantly not expressed for small $\alpha_I$. However, the range over which the switch occurs is different. In  Figure \ref{fig:Jonas_1}C we plot the  fraction of time the simulation shows the reporter expressed (which corresponds to the fraction of cells that are fluorescent in a mother machine) as a function of the relative synthesis rate of SinI and SinR. In the following, this quantity will be recalled as Matrix-Producing Fraction (MPF).
With the larger absolute numbers of proteins, the switch is much more sensitive to the relative synthesis rates of the two competing proteins, as shown by the different slopes of the curves in  Figure \ref{fig:Jonas_1}C. In addition, there is a shift of the curves to the left (towards smaller SinI synthesis rates) with increasing stochasticity, most pronouncedly when comparing the small and large protein copy number situation, but also when increasing stochasticity by increasing burst sizes. Thus stochastic bursts of synthesis of SinI allow for the expression of matrix genes even when the SinI synthesis rate is smaller than that of SinR.

\begin{figure}[tb]
\centering
\subfigure{\includegraphics[width=1\textwidth]{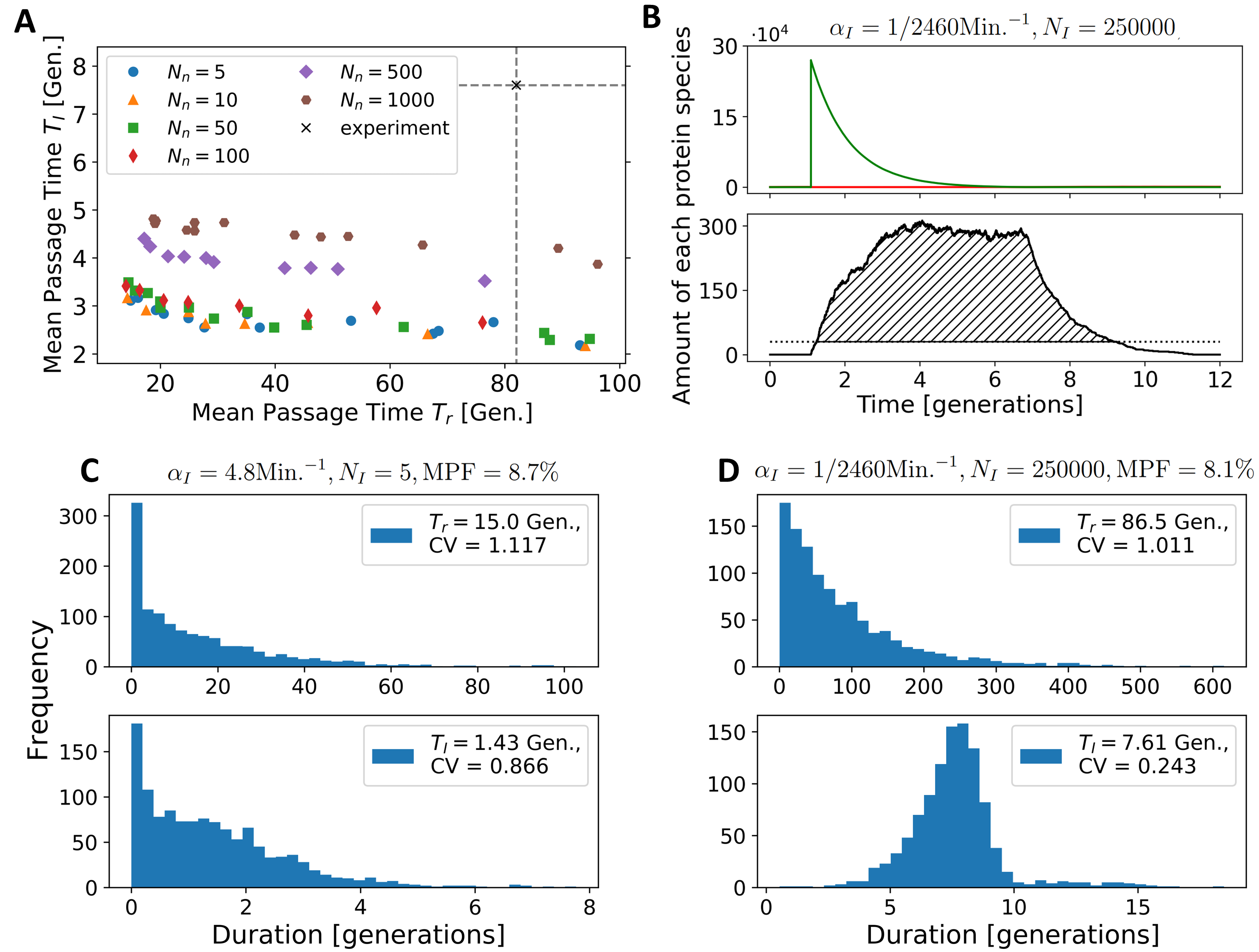}}
\caption{Lifetimes of the phenotypic states in the SinI-SinR competition model: \textbf{A} Durations $T_r$ and $T_I$ of the motile and matrix-producing state for different effective synthesis rate $N_I\alpha_I$ with varying the burst size $N_I$ of SinI. Experimental values of the durations are indicated by the black cross and require unrealistically high bursts. \textbf{B} Stochastic time evolution of the concentrations of SinR (red), SinI (green) and the matrix reporter protein (black) after such a very large SinI burst, showing prolonged expression of the matrix reporter.  \textbf{C}, \textbf{D} Distributions of the durations of the motility state (top) and of the matrix-producing state (bottom) for two parameter sets with approximately the same matrix-producing fraction and without (\textbf{C}) or with (\textbf{D}) unrealistically large bursts. }
\label{fig:Jonas_2}
\end{figure}

For a quantitative comparison with experimental results, we consider the lifetimes of the motile and matrix-producing states. Norman  et al. \cite{Norman} observed that the motile state is rather stable with an average lifetime of 81 generations, while the matrix-producing state is rather short-lived with an average lifetime of 7.6 generations. Moreover, the two distributions of the lifetimes are different: The lifetime of the motile state is exponentially distributed, indicating a state without memory, and exhibits a coefficient of variation of $\text{CV} \approx 1$, whereas the duration of the matrix-producing state is pronouncedly non-exponential with a maximum at a finite value and  $\text{CV} < 1$. As seen before in Fig. \ref{fig:Jonas_1}, the minimal stochastic competition model reproduces these features qualitatively by matching the MPF. There are, however, qualitative discrepancies; specifically, the model underestimates the duration of the matrix-producing state (Figure  \ref{fig:Jonas_2}A). This value cannot easily be adjusted by modifying model parameters. While the duration of the motile state can be adjusted by modifying the total synthesis rate of SinI, the duration of the matrix state is seen to depend on the size of SinI bursts, which is visualized in \ref{fig:Jonas_2}A. To obtain a duration of 7.6 generations, extremely large bursts, exceeding a burst size of $N_I=1000$ and thus far outside the typical range \cite{taniguchi2010quantifying} are required.
Figure \ref{fig:Jonas_2}B shows the dynamics of a transition induced by such a huge burst of SinI. In that scenario, the duration of the matrix-producing state is prolonged by the slow degradation of the unrealistically large number of SinI molecules. In Figures \ref{fig:Jonas_2}C and D, distributions of the lifetimes of both state are shown for two different parameter sets that results in approximately the same MPF's, one with realistic synthesis rates and burst sizes (Figure \ref{fig:Jonas_2}C) and one for very large bursts (Figure \ref{fig:Jonas_2}D). The motility state is indeed memoryless (with a CV$\approx 1$), independent of the choice of parameters, whereas the matrix-producing state shows a distribution similar to the experimental one (with mean 7.6 generations and a low CV) only for very large bursts, but a more exponential distribution with a smaller mean for parameters in the realistic range.

All these observations indicate that intrinsic fluctuations within such stochastic competition are unlikely to provide the switching mechanism. Rather a source of fluctuations that is extrinsic to the SinR-SinI-SlrR core circuit is needed to generate a sufficiently strong SinI pulse and induce switching.

\subsection{Switching triggered by  Spo0A-P fluctuations}

In the cellular context, SinI is controlled by the phosporylated form of Spo0A (Spo0A-P) and induced as part of the starvation stress response. The phosphorelay which regulates the activity of Spo0A is a known source of stochasticity \cite{Losick2017} and also governs entry into sporulation via periodic peaks in the Spo0A-P concentration under starvation \cite{Schultz, Igoshin1, Igoshin2}. We therefore ask whether fluctuating levels of Spo0A-P can provide the stochastic input needed to trigger SinI synthesis and the switch to matrix production under the steady state conditions studied here.
 
\begin{figure}
    \centering
    \includegraphics[scale = 0.16]{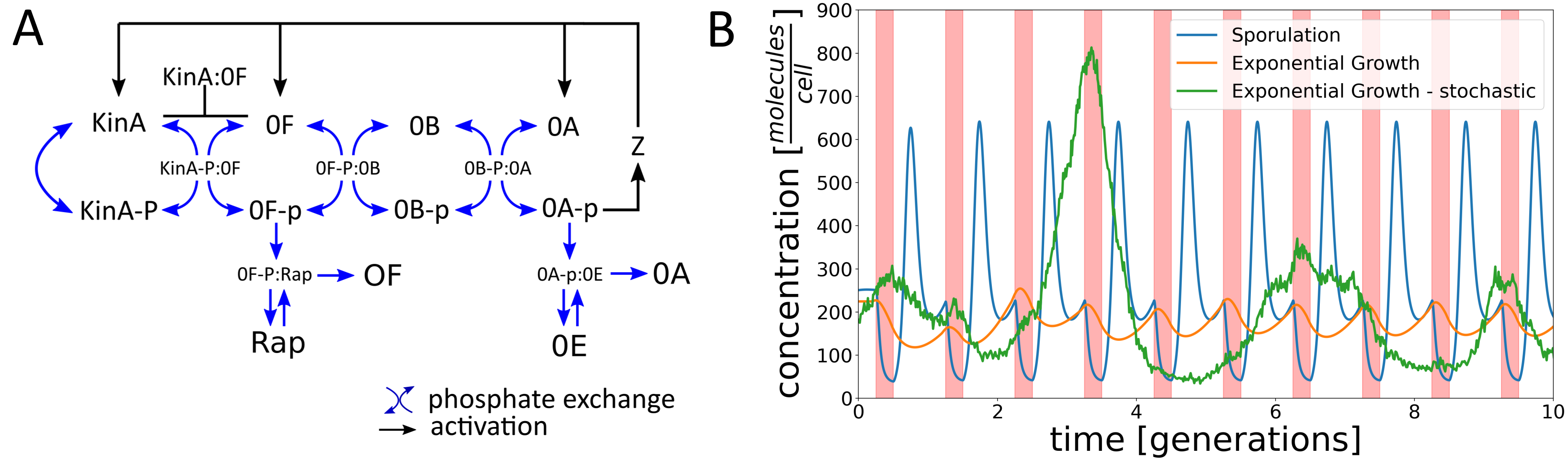}
    \caption{ Model of the phosphorelay for the activation of Spo0A: \textbf{A} Schematic diagram of the phosphorelay 
    that shows the tranfer of a phosphate group between KinA, Spo0F, Spo0B and Spo0A (blue arrows) and the transcriptional feedback by activation of KinA, Spo0F, and Spo0A induced by Spo0A-P (drawn after Ref. \cite{Igoshin2}). 
    \textbf{B} Dynamics of the Spo0A-P  concentration development for different growth conditions (sporulation conditions vs. growth) and different models (deterministic vs. stochastic). Note that the time is given in generations to facilitate comparison of different conditions. }
    \label{Network_Spo0Ap}
\end{figure}

Narula et al. showed that the concentration of Spo0A-P undergoes cyclic peaks when triggering entry to sporulation \cite{Igoshin2}. They attributed the emergence of the peaks to the architecture of the complex phosphorelay in which phosphate is  transferred to Spo0A via several intermediate steps (Figure \ref{Network_Spo0Ap}A)
 and identified three key features of the phosprelay essential for the Spo0A-P peaks under sporulation: 1. the negative autoregulatory feedback of Spo0A-P caused by the complex formation of Spo0F and KinA; 2. a synthesis imbalance due to temporal changes in the Spo0F and KinA gene dosage during DNA replication, as the \emph{spo0F} gene is located close to the origin of replication, while the \emph{kinA} gene is closer to the replication terminus; and 3. a delayed feedback loop between phosphorylation of Spo0A and synthesis of the relays' components \cite{Igoshin2} . 
 
We transferred their model to our conditions, in which cells grow faster, by using the assumption that the relevant  protein concentrations  in the relay stay the same compared to sporulation. Time series of the Spo0A-P concentration under this condition are shown in Figure \ref{Network_Spo0Ap}B. We find that a deterministic model does not show the strong regular pulses in the concentration of Spo0A-P as seen in the initiation of sporulation (blue curve in Fig. \ref{Network_Spo0Ap}B), but rather oscillations with a small amplitude (orange curve in Fig. \ref{Network_Spo0Ap}B).


However, this behavior is altered dramatically when a stochastic variant of the model is used that includes number fluctuations and bursty protein production (green curve in Figure \ref{Network_Spo0Ap}B). In this case, the amplitude of the oscillations displays large fluctuations that can span over several periods of the deterministic oscillation. Their maxima coincide with periods of imbalance in gene abundance. These fluctuations result in irregular pulses of Spo0A-P with peak concentrations similar to  the sporulation case.
Hence, if the cell  maintains identical steady state concentrations under faster growth, large fluctuations of the Spo0A-P concentration could be accomplished as well.

\begin{figure}
    \centering
    \includegraphics[scale = 0.18]{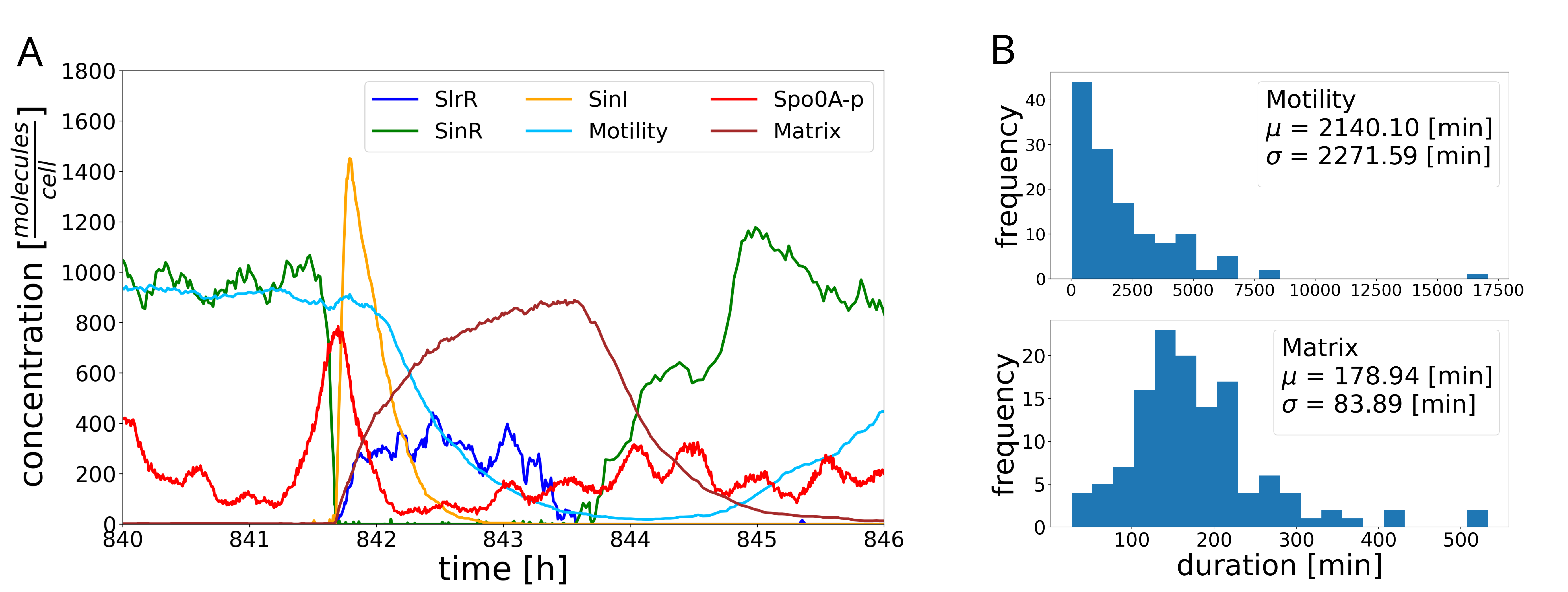}
    \caption{Induction of the switch to matrix production by fluctuations in the phosphorelay: \textbf{A} Time evolution of the concentrations of proteins  in the SinR-SlrR-SinI network including Spo0A-P. The fluctuating concentration of Spo0A-P is determined by the phosphorelay and coupled to SinI via an activation function. \textbf{B} Histograms of the lifetimes of the resulting matrix and motility states. }
    \label{Spo0A-p_trajectory}
\end{figure}
We next tested whether these pulses in the Spo0A-P concentration work as a trigger for matrix production, since Spo0A-P is known to be an activator of SinI. Hence, we coupled the stochastic model of the phosphorelay to the SinR-SlrR-SinI network by making the SinI synthesis rate dependent on the concentration of Spo0A-P (see Methods). We then simulated the dynamics of the SinR-SlrR-SinI network as driven by the fluctuations in the phosphorelay. A corresponding trajectory is plotted in Figure \ref{Spo0A-p_trajectory}A. Here, it can be seen that pulses in Spo0A-P (red curve) can indeed induce a SinI pulse (yellow curve) and trigger a transient switch to matrix production (biofilm reporter in brown) similar to what was seen with an externally provided SinI pulse in Figure \ref{involving_SinI}. These switches between lifestyles can be observed multiple times throughout the entire trajectory. We quantified them by creating histograms of the individual lifetimes of each state in Figure \ref{Spo0A-p_trajectory}B.

In these histograms a ratio of  of 1.06 between the mean and standard deviation (CV) is achieved for the motility state and of 0.46 for the biofilm state, in agreement with the exponential nature of entry in the biofilm state and precisely timed duration of staying in it.
Thereby, our results suggest that fluctuations in the Spo0A-P phosphorelay trigger the switch to matrix production in \textit{Bacillus subtilis}.
\section{Discussion}

In this study, we have analyzed theoretical models for the phenotype switch between motility and matrix production in {\it B. subtilis}, the first step towards the formation of a biofilm. The core regulatory circuit, the SinR-SlrR-SinI system, consists of two modules, a bistable switch based on mutual negative control of SinR and SlrR, and a sequestration module based on the binding of SinR and SinI. The mutual negative control of SinR and SlrR resembles other bistable switches \cite{Veening}, but contrary to the paradigmatic design of mutual transcriptional repression (the "toggle switch"), the design of the SinR-SlrR switch is asymmetric in the sense that SinR represses SlrR transcriptionally, while SlrR inactivates SinR by sequestration. This design is notable from a theoretical point of view, as sequestration provides a strongly nonlinear dependence on concentration, allowing for ultrasensitive response functions \cite{buchler2009protein}. Our results indicate that such nonlinearity would allow for bistability in an asymmetric repression-sequestration switch even in the absence of cooperative transcription factor binding, while cooperativity is known to be essential in symmetric switch designs based on mutual transcriptional repression. However, the SinR-SlrR system does not make use of this scenario.  

The deterministic variant of our model for the SinR-SlrR switch shows that a pulse of SinI (which may be triggered by upstream signals via phosphorylated Spo0A) induces a transition from the motile state to the matrix-producing state by shifting the location of the bistable region in the parameter space. Depending on the values of the (unrepressed) synthesis rates of SinR and SlrR, the switch may be reversible or irreversible: If cells observed to be motile are motile in a monostable region of the parameter space, a SinI pulse drives these cells into the monostable matrix-producing state and back to the monostable motile state. In this case, the cells go through a deterministically prescribed dynamic program that is driven by the dynamics of the SinI concentration. If, however, the initially motile cells are taken as motile from the bistable region of the parameter space, a pulse of SinI only drives them into the matrix-producing state and when the pulse has vanished, the cells do not return to the motile state but continue to produce matrix until eventually a stochastic transition takes place. The experiments of Norman et al. \cite{Norman} suggest the former scenario, as the transitions from the matrix-producing state to the motile state show a history dependence or "memory", in the sense that the observed transition rate increases with time in the matrix-producing state. However, this interpretation is complicated due to the huge difference in stability between the two phenotypic states. Compared to the long lifetime of the motile state, transitions are indeed very fast and the duration of the transitions can be neglected, but the transition duration does possibly contribute to the memory seen for the matrix-producing state. Thus the second scenario with stochastic return to the motile state cannot fully be ruled out yet. Conceptually, the two scenarios are quite distinct, with motility as the single stable state and only transient dynamic expression of matrix genes in the first case and two stable states  with stochastic switching between them in the second. In practice, the observable difference between the two scenarios is however expected to be small, again due to the short lifetime of the matrix-producing state \cite{Norman}.  

If the return to motility is stochastic as well, the question arises how the switch back from the matrix-producing state to the motile state is triggered. Based on our stochastic modeling, both states are rather stable, so that intrinsic fluctuations in the synthesis of SinR and SlrR appear unlikely to be sufficient to induce the switch. Therefore, just like SinI triggers the switch from motility to matrix production, we suspect that another factor may be involved in the switch from matrix production to motility, possibly by controlling the halflife of the SlrR protein \cite{Chai2010b,newinsights,Vlamakis}. Several factors have indeed been reported that decrease the fraction of matrix-producing cells in a population and might be candidates for such a regulator, including DegU \cite{Marlow2014} and YmdB \cite{kampf2018selective}.

Switching into the matrix-producing state under biofilm-forming conditions is induced by upstream signaling via the phosphorylation of Spo0A, which in turn is controlled by a phosphorelay consisting of Spo0F, Spo0B, Spo0A and the histidine kinases KinA, KinB, KinC, and KinD \cite{Burbulys1991,Tzeng1998} that can integrate different extracellular and intracellular signals \cite{arnaouteli2021bacillus}. However, switching is also observed under conditions of steady state growth in the absence of stress signals, where it is triggered by stochastic fluctuations rather than explicit signals \cite{Norman}. While it is clear that under these conditions some stochastic trigger is at work, it is less clear in which part of the circuit that stochasticity is generated. Here we have analyzed several possible sources of noise using stochastic models of the genetic circuit. Our models suggest that the fluctuations responsible for the switch are not generated in the switch itself, but rather transmitted from the upstream signaling pathway, i.e., from Spo0A and the phosphorelay. Specifically, our model suggests that intrinsic fluctuations in the SinI-SinR competition, while in principle able to trigger switching, cannot explain switching quantitatively. Unless the noise level is artificially increased, e.g. by huge burst sizes in protein synthesis, intrinsic noise alone underestimates the stability of the matrix-producing state. Instead, a source of noise extrinsic to the SinR-SinI-SlrR core circuit generates a pulse of SinI, followed by almost deterministic dynamics. Our analysis of the coupling to the phosphorelay indicates that this can indeed be achieved by fluctuations arising in the phosphorelay and transmitted to SinI via phosphorylated Spo0A.

Under conditions in which a biofilm is formed, the growth of the cells slows down during the switch to matrix production \cite{Igoshin2023}, in contrast to the constant conditions in a mother machine or other microfluidic setups \cite{Norman,kampf2018selective}. Slowing growth affects the expression of all genes through reduced dilution, but also via the availability of gene expression machinery \cite{Klumpp2009}. Growth rate changes can also affect the expression of different genes differentially, e.g. if their protein products have different lifetimes. A recent study \cite{Igoshin2023} has proposed, based on a model supported by expression time series in various mutants, that the growth reduction provides an additional incoherent feedforward effect on matrix production genes: A growth reduction resulting from moderate starvation activates SinI via the phosphorylation of Spo0A, but indirectly shifts the balance between SinI and SlrR, effectively stopping matrix production. As a result, strongly starved cells do not produce matrix despite the phosphorylation of Spo0A, but rather sporulate, an alternative stress response, mutually exclusive with matrix production \cite{Igoshin2023}. This second phenotypic decision thus does not induce a reverse transition, even though matrix production will stop, but rather further differentiation that results in additional heterogeneity in the forming biofilm that is not present during the stochastic transitions in approximately constant growth conditions. 

In summary, we have studied here several variants of a model for the genetic circuit underlying the phenotype switch from motility to matrix production, an early step in the production of a biofilm. Specifically, we discussed the roles of the different sub-modules of the circuit and considered stochastic phenotype switching as seen without external triggers such as starvation. However, even under conditions of starvation, stochasticity is crucial as stochasticity allows only a subpopulation of the cells to induce the transition, thus enabling the differentiation seen in the biofilm. In general, the models studied here and in previous work \cite{Norman, Paulsson2019, Igoshin2023} suggest that the decisions about phenotype switching or cell fate are made by an interplay of dedicated dynamic expression programs, different sources of stochasticity and the overall physiological state of the cells as reflected in their growth state to allow dynamic responses to stress as well as  the development of a community of differentiated cells.

\begin{acknowledgments}
The authors thank Jan Kampf and J\"org St\"ulke for discussions and Oleg Igoshin for communicating their recent results \cite{Igoshin2023} before publication. The simulations were run on the GoeGrid cluster at the University of G\"ottingen, which is supported by the Deutsche Forschungsgemeinschaft (DFG, German Research Foundation) – project IDs 436382789; 493420525 and MWK Niedersachsen (grant no.\ 45-10-19-F-02).
\end{acknowledgments}
\section{Methods}

\subsection{Deterministic model for the SinR-SlrR-SinI network}
\label{methods_SlrRSinR_deterministic}
The time evolution of the protein concentrations in  the SinR-SlrR-SinI network (without Spo0A) as shown in Figure \ref{network_bistabiltiy}A is calculated with the following set of equations.
\begin{equation}
    \begin{split}
        \Dot{s}  &= \alpha_s R(r) - k_+ s r - \beta s \\
        \Dot{r}  &= \alpha_r  - k_+ s r  - k_+ r I - \beta r \\
        \Dot{I}  &= \alpha_I   - k_+ r I  - \beta_I \\
        \Dot{c}  &= k_+ s r - \beta c 
    \end{split}
\label{network}
\end{equation}
Here $s$, $r$, $I$ and $c$ are the concentration of SlrR, SinR, SinI and the SlrR-SinR complex, respectively. The synthesis rate of SlrR is described by $\alpha_s R(r)$, incorporating repression by SinR, and the synthesis of SinR and SinI are given by the constants $\alpha_r$ and $\alpha_I$. Here, $\alpha_{s,r,I}$ are the respective synthesis rates and $R(r)$ is a Hill function $R(r) = \frac{1}{1 + (r/K_r)^{n_r}}$, in which $K_r$ is the affinity and $n_{r}$ the cooperativity of the binding. The complex formation of SinR and SlrR and SinI and SinR follow simple reaction kinetics and can be written as $k_+ s r$ and $k_+ r I$, respectively. Due to the high binding affinity in both cases, complex formation is taken to be irreversible. All molecules are reported to be stable and hence, we regarded dilution by growth as the main paths of degradation (for SlrR a half life of 100 min has been reported , which is likely crucial under starvation conditions, but long compared to the doubling time under the conditions studied here  \cite{Chai2010b}). Dilution is described by a simple decay process with constant rate $\beta$ which is the inverse of the cell cycle duration. The reporter for motility $G_{Mo}$  and Biofilm formation $G_{Bio}$ are downstream of this network and their concentrations are calculated via $\Dot{G}_{Mo} = \alpha_m R(c) - \beta G_{Mo}$ and  $\Dot{G}_{Bio} = \alpha_m R(r) - \beta G_{Bio}$. An overview of the parameters and their values can be found in Table \ref{parameter_overview}.

To determine the bifurcation points in Figure \ref{network_bistabiltiy}B and \ref{involving_SinI}C we determine the steady state of equation \ref{network} and simultaneously set one of the eigenvalues of the corresponding Jacobian matrix (chosen by biological constraints on the parameters) to zero. This was done for a given value of $\alpha_s$ and gave us five equations to determine the four concentrations and $\alpha_r$. The resulting combinations of $\alpha_r$ and $\alpha_s$ determine the boundaries of the bistable region.  The steady state concentrations for all combinations of $\alpha_r$ and $\alpha_s$ were calculated by setting equation \ref{network} to zero using a Python rootfinding algorithm.

\subsection{Stochastic simulations of the switch}
\label{methods_SlrRSinR_stochastic}
To model stochastic fluctuations in the protein concentrations, simulations with the Gillespie algorithm are used \cite{Gillespie}. 
In these simulations, the current (microscopic) state of the system is described by the copy numbers of all protein species. In each step of the algorithm transitions between such states are carried out by changing the copy numbers of a certain species due to synthesis, degradation and complex formation.  All possible transitions are listed in table \ref{parameter_overview}. To incorporate burst in protein synthesis, the number of proteins created in one synthesis step is drawn from a geometric distribution with a mean burst size $N_b = 10$ analogous to earlier work \cite{Norman}. This procedure allows us to model individual trajectories of the protein concentrations. In contrast to the deterministic rate equations, we can in principle also see transitions between steady states, caused by stochastic fluctuations which destabilize the system.


\subsection{Minimal competition model}
\label{subsec:methods_Jonas}
The same procedure is used to describe and simulate the minimal SinI-SinR competition model as introduced  in \cite{Paulsson2019}. This model contains only the synthesis rates $\alpha_r$ and $\alpha_I$ of SinR and SinI, the complex formation rate $k_+$ and the dilution rate $\beta$. The matrix reporter is synthesised with a rate $\alpha_z\cdot R(r)$. To rule out that small fluctuations can causing Biofilm initiation 
When the reporter copy number exceeds a  threshold of $0.1\tau\alpha_I$, matrix production is considered as active. This threshold makes sure that small fluctuations of matrix reporter do not get counted as transitions to the matrix production state. It is set to $0.1\tau\alpha_I$ which here corresponds to approximately 30 proteins and denotes 10\% of the mean maximal SinI concentration. Moreover, 
synthesis of SinR and SinI occurs in bursts with average burst size of $N_r = N_n = 5$ unless stated otherwise.

The behaviour of the system is characterized by four key quantities: the Biofilm fraction (BF) that denotes the fraction of time spent in the matrix production state, the mean passage times (MPT) for each state ($T_r$ for motility, $T_I$ for matrix production) and the standard deviation (CV) of the MPT distribution.

\subsection{Model of the Spo0A phosphorelay}
\label{methods_Spo0A}

The model for the dynamics of the Spo0A phosphorelay is based on work by Narula et al. \cite{Igoshin2}. The equations and rate constants used here are as in their paper and can in principle be constructed in the same manner as the rate equations for the network of SlrR-SinR-SinI by using similar terms to describe the processes of synthesis, complex formation and additional Hill functions for phosphorylation/dephosphorylation. Also the transfer of the rate equations into a master equation for the stochastic version of the model is done in a completely analogous fashion to the case of SinI-SinR-SlrR. Protein bursts are again modeled by drawing the number of new molecules from a geometric distribution.

We adapted the model to faster growth under our conditions by changing the degradation rate according to the faster dilution due to cell growth and assured the same steady state concentrations by rescaling the synthesis rates with the ratio of the dilution rates $k_{\text{fast}}/k_{\text{slow}}$ in which $k_i$ is the degradation rate of fast or slow growth. The period of gene copy number imbalance (where the \emph{spo0F} gene is already replicated, but not \emph{kinA}) was set to a quarter of the cell cycle.

The phosphorelay is coupled to the SinI-SlrR-SinR network by making the synthesis rate of SinI $\alpha_I$ dependent on the concentration of phosphorylated Spo0A via an activating Hill function, 
\begin{align}
    \alpha_I =\alpha_I^0   \frac{(0A/K_{0A})^{n_{0A}}}{1 +(0A/K_{0A})^{n_{0A}} }.
\end{align}
Here $w$ describes the fold change in increase once sufficient Spo0A-P is present.

\begin{table}[H]
\begin{tabular}{|c||c|c||c|c||c|c|}
\hline
    \multicolumn{3}{|c||}{}     & \multicolumn{2}{c||}{synthesis}         & \multicolumn{2}{c|}{degradation} \\ 
\hline
index & protein/dimer   & abbr. & transition          & rate            & transition          & rate      \\
\hline
1     & SinR            & $r$   & $r \rightarrow r+1$ & $\alpha_r$      & $r \rightarrow r-1$ & $\beta r$ \\
\hline
2     & SlrR            & $s$   & $s \rightarrow s+1$ & $\alpha_s R(r)$ & $s \rightarrow s-1$ & $\beta s$ \\
\hline
3     & SinI            & $I$   & $I \rightarrow I+1$ & $\alpha_I$      & $I \rightarrow I-1$ & $\beta I$ \\
\hline
4     & SinR-SlrR-Dimer & $c$   & \begin{tabular}[c]{@{}l@{}}$c \rightarrow c+1$\\ $r \rightarrow r-1$\\ $s \rightarrow s-1$\end{tabular} & $k_c sr$        & $c \rightarrow c-1$ & $\beta c$ \\
\hline
5     & SinR-SinI-Dimer & $d$   &    \begin{tabular}[c]{@{}l@{}}$d \rightarrow d+1$\\ $r \rightarrow r-1$\\ $I \rightarrow I-1$\end{tabular}   & $k_d rI$        & $d \rightarrow d-1$ & $\beta d$ \\
\hline
6     & Motility Reporter & $Y$   & $Y \rightarrow Y+1$ & $\alpha_Y R(c)$      & $Y \rightarrow Y-1$ & $\beta Y$ \\
\hline
7     & Matrix Reporter & $Z$   & $Z \rightarrow Z+1$ & $\alpha_Z R(r)$      & $Z \rightarrow Z-1$ & $\beta Z$ \\
\hline
\end{tabular}
\caption{List of transitions and corresponding rates of the stochastic model}
\label{transition_overview}
\end{table}

\begin{table}[H]
\centering
\begin{tabular}{|c|c||c|c|}
\hline
parameter & value   & parameter & value      \\
\hline
$\beta$ & 1/30 $\text{min}^{-1}$   & $k_+$ & 1000  $\text{min}^{-1} \, (\text{molecules/cell})^{-1}$ \\
$K_r$ & $50\,  \text{molecules/cell}$   & $n_r$ & 4   \\
$\alpha_m$ & 30  $\text{min}^{-1}$   & $K_{0A}$  & 800 $\text{molecules/cell}$  \\
$\alpha_I^0$ & 75  $\text{min}^{-1}$   & $n_{0A}$  & 8   \\
\hline

\hline
\end{tabular}
\caption{Standard values used for SinR-SlrR-SinI network calculations and for the coupling of SinI to Spo0A-P}
\label{parameter_overview}
\end{table}


%


\section{Supplements}

\subsection{Analytical Approximation}
\label{methods_analytical_approx}
For an analytical approximation of the time in which SinR is not present after a SinI pulse, we separate the process during and after a pulse of SinI synthesis into three phases. The first phase represents the period in which SinR is actively titrated by newly produced SinI. In the second phase, pools of SinI and SlrR are built up due to ongoing SinI synthesis after SinR is fully titrated. These pools are then depleted in the third phase when SinI synthesis is turned off. Therefore, the total time in which no SinR is present results from adding the durations of the second and third phase. 

We first calculate the time it takes until the SinR concentration drops to zero. This will later give us the duration of the second phase in which SinI and SlrR production is unperturbed.
Since SinR is still present and SlrR repression is not lifted yet, because newly synthesised SinI and SlrR proteins are immediately bound in complexes with SinR, so $s = I = 0$. Furthermore, the two concentration stay zero for the considered time period and hence their time derivative is also zero. This leads to modified rate equations given by $\Dot{s}  = - k_+ s r =0 $, $\Dot{I} = \alpha_I - k_+ I r = 0$ and $\Dot{r}  = \alpha_r - \beta r - k_+ s r - k_+ I r$. 
Here and in the following, the repression function $R(r)$ is approximated by a step function, $R(r)=1$ for $r<K$ and $R(r)=0$ for $r\geq K$. 

Plugging everything into the time derivative of $r$ leads to $\Dot{r}  = \alpha_r - \beta r -  \alpha_I$.
With the initial condition $r(0) = \alpha_r/\beta$, the solution is
\begin{equation}
    r(t) = \frac{\alpha_I}{\beta} e^{-\beta t} - \frac{\alpha_I - \alpha_r}{\beta}.
\end{equation}
Hence we can obtain the time needed until all SinR molecules are titrated as 
\begin{equation}
    t_1 = \frac{1}{\beta} \text{ln}\big(\frac{\alpha_I}{\alpha_I - \alpha_r}\big)
\end{equation}
This result allows us to calculate the time in which a pool of SinI and SlrR can be built up that protects against repression. 
To calculate how many proteins are produced in that time,  we introduce the new variable $Z = I + s$.   
Simplifying the rate equations for the second phase (with $r \approx 0$) results in 
$\Dot{r} = - k_+(Ir - sr) + \alpha_r =0 $ and $\Dot{Z} = -\beta Z - \alpha_r + \alpha_s + \alpha_I$.  
With $Z(t_1) = 0$, this leads to
\begin{equation}
    Z(t) = \frac{\alpha_s + \alpha_I - \alpha_r}{\beta}\big( 1 - e^{-\beta (t - t_1)}\big).
\end{equation}
The pool of SinI and SlrR proteins which need to be titrated by SinR after the end of the SinI pulse is then
\begin{equation}
    Z(t_{\text{on}}) = \frac{\alpha_s + \alpha_I - \alpha_r}{\beta}\big[1 - e^{\beta t_{\text{on}}} (\frac{\alpha_I}{\alpha_I - \alpha_r})\big]
\end{equation}
with $t_{on}$ being the duration of the SinI pulse.

At last, the time is calculated that is needed to titrate this pool after the SinI pulse. In this third phase, the dynamics is the same as in the second phase, 
but with $\alpha_I = 0$ and initial conditions $Z(t_{\text{on}})$. This results in 
\begin{equation}
    Z(t) = \big(Z(t_{\text{on}}) + \frac{\alpha_r - \alpha_s}{\beta}\big) e^{-\beta (t - t_{\text{on}})} - \frac{\alpha_r - \alpha_s}{\beta}
\end{equation}
From this, the time $\Delta t$, in which no SinR is present, is obtained by solving for $Z(t_2) = 0$ and subtracting $t_1$, 
\begin{equation}
    \Delta t = t_2-t_1 
    = \frac{1}{\beta} \, \ln\bigg(\frac{\alpha_I - \alpha_r}{\alpha_r - \alpha_s} e^{\beta t_{\text{active}}} - \frac{\alpha_I}{\alpha_r - \alpha_s} +1\bigg).
\end{equation}
\clearpage

\clearpage

\begin{figure}[!h]
    \centering
    \includegraphics[scale = 0.5]{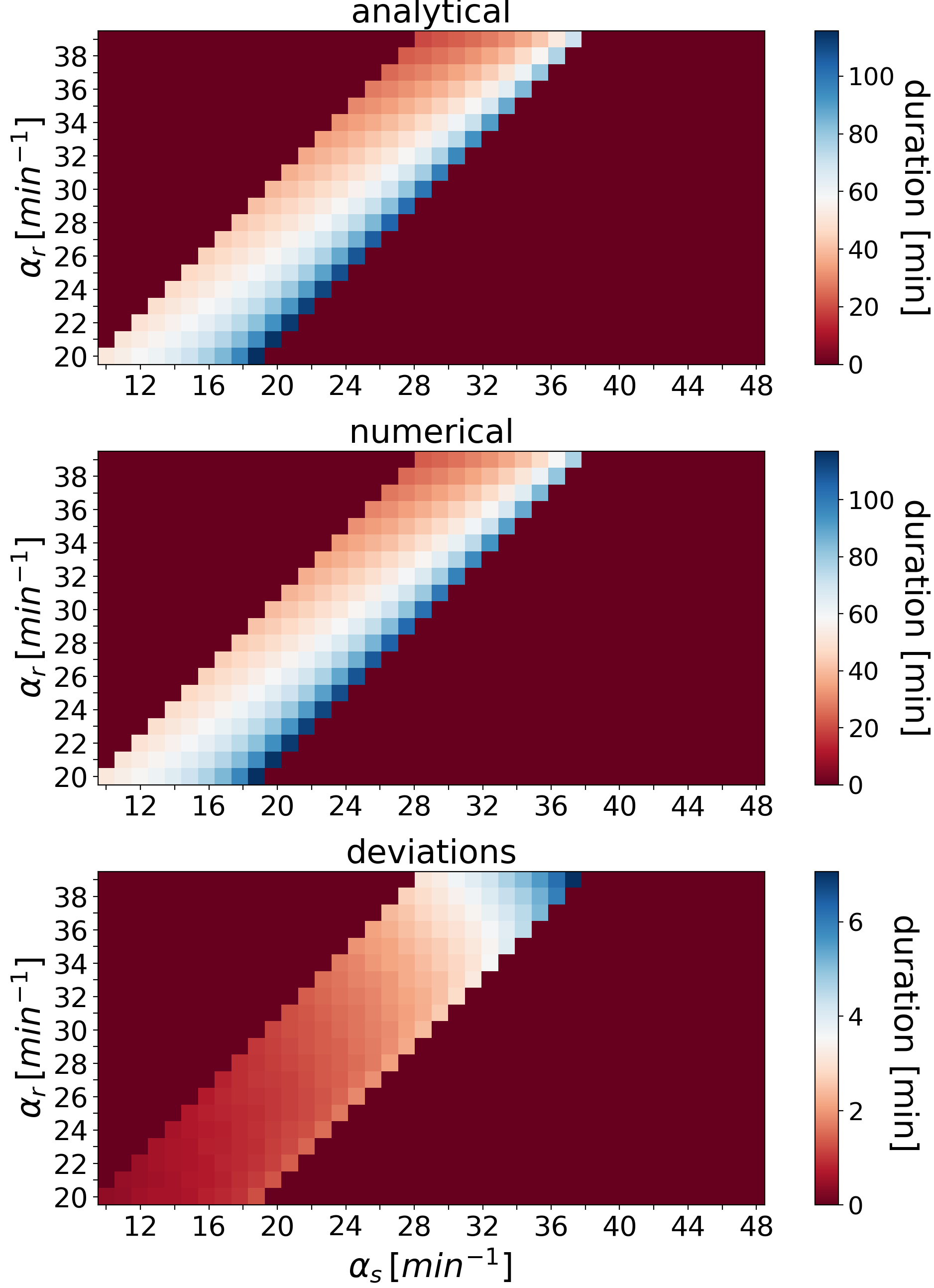}
    \caption{Duration how long the SinR concentration stays 0 after a SinI pulse. In the first row analytical and in the second numerical results are shown. The dark red region corresponds to unexplored parameter space. The last row shows the deviation of the two solutions.}
    \label{analytical_comp_supplements}
\end{figure}

\clearpage

\end{document}